\documentstyle[multicol,pre,aps,epsfig]{revtex}
\begin{document}
\draft
\title{Etched distributed Bragg reflectors as three-dimensional
photonic crystals: photonic bands and density of states}
\author{E. Pavarini and L.C. Andreani}
\address{ Istituto Nazionale per la Fisica della Materia
and Dipartimento di Fisica ``A. Volta",
Universit\`a di Pavia, Via Bassi 6, I-27100 Pavia, Italy
}
\date{\today}
\maketitle

\begin{abstract}
The photonic band dispersion and density of states (DOS)
are calculated for the three-dimensional (3D) hexagonal structure 
corresponding to a distributed Bragg reflector 
patterned with a 2D triangular lattice of circular holes.
Results for the Si/SiO$_2$ and GaAs/AlGaAs systems determine 
the optimal parameters for which a gap in the 2D plane occurs 
and overlaps the 1D gap of the multilayer. 
The DOS is considerably reduced in correspondence 
with the overlap of 2D and 1D gaps.
Also, the local density of states (i.e., the DOS weighted
with the squared electric field at a given point)
has strong variations depending on the position.
Both results imply substantial changes of spontaneous emission
rates and patterns for a local emitter embedded in the structure
and make this system attractive for the fabrication 
of a 3D photonic crystal with controlled radiative properties.
\end{abstract}
\vskip 1cm

\begin{multicols}{2}
\section{Introduction}
\label{sec:intro}

Photonic crystals are being intensively studied with the goal 
of achieving a photonic band gap as well as control of light 
propagation and light-matter interaction at visible 
frequencies \cite{yablonovitch87,john87,joannopoulos_book,sakoda_book}.
A complete photonic gap in three dimensions (3D) has been 
demonstrated for the  diamond lattice of dielectric spheres \cite{ho90}, 
the fcc lattice of air spheres 
(or ``inverse opal'') \cite{sozuer92,busch98}, 
the ``yablonovite'' \cite{yablonovitch91}, 
the ``woodpile'' \cite{ho94},
and other more complex 3D geometries \cite{fan94,johnson00}.
All these structures are difficult to fabricate at optical wavelengths:
they often require the use of a bottom-up procedure,
like e.g. in the case of inverse opals, where the template
is first built by self-assembling of dielectric spheres in a solution 
and the voids are subsequently filled with a high refractive index 
material \cite{thijssen99,blanco00,vlasov01}.
As an alternative, two-dimensional (2D) photonic crystals can be 
fabricated with a top-down approach based on lithography and etching.
This procedure can be realized at sub-micrometric wavelength scales 
and allows for the controlled introduction of linear and point defects.
A main problem however concerns the control of light propagation 
in the third (vertical) dimension: even in the case of a
waveguide-embedded 2D photonic crystal, radiation losses due to
out-of-plane diffraction in the vertical direction cannot be eliminated
for photonic modes which lie above the light 
line \cite{gourley94,russell96,benisty99}.

In this work we explore theoretically another possibility
for achieving control of light propagation in 3D, namely
the use of distributed Bragg reflectors (DBRs) patterned
with a 2D lattice in the layer planes. DBRs are dielectric multilayers 
which are periodic in one dimension (1D) \cite{born-wolf,yariv-yeh,mcleod},
i.e., they represent 1D photonic crystals and have a band gap 
for propagation of light along the multilayer axis. By patterning and
etching a DBR with a 2D lattice which possesses a photonic gap,
a 3D photonic crystal with uniaxial (or biaxial) symmetry is obtained:
if the 1D gap of the multilayer is made to coincide with
the gap of the 2D structure, a common band gap for light propagating
along the main crystal axes can be achieved.
Although a complete photonic band gap in 3D is not expected 
(due to the angular dependence of the 1D and 2D gaps, 
or else to the anisotropy of the Brillouin zone),
an appreciable control of the photonic density of states (DOS) and 
consequently of the spontaneous emission properties may be realized.
In such a situation, the etched DBR represents a favorable system
for the study of spontaneous emission control in 3D photonic systems,
since local (and possibly active) defects can be introduced 
at the level of 1D epitaxial growth and of 2D lithographic design.

Here we focus on the simplest and most promising structure,
namely a DBR patterned with a triangular lattice of holes:
the planar structure is known to display a 2D photonic gap 
common to both polarizations of light, if the dielectric 
constant and air filling fractions are large 
enough \cite{joannopoulos_book,villeneuve92,meade92,padjen94}.
We calculate the 3D photonic band dispersion and the photonic DOS
for parameters which are representative of the Si/SiO$_2$ system
(a typical high-index contrast DBR) and of the GaAs/Al$_x$Ga$_{1-x}$As
system (with low index contrast). In particular, we determine the 
conditions for the 1D and 2D photonic gaps to overlap: since the 
1D stop band of the DBR changes upon patterning, and the 2D 
lattice of holes is formed in a stratified medium, the overlap 
of the two band gaps represents a sort of selfconsistency problem. 
The density of states is found to be considerably reduced
in correspondence with the overlapping 2D and 1D gaps.
Also, the local density of states (which is the relevant
quantity for spontaneous emission rates) shows strong variations
and additional reductions as a function of position.
These findings show that sizeable changes of spontaneous emission rates 
and patterns of a local emitter embedded in this anisotropic 
3D structure should occur.
The present results may also serve as guidelines for experimental 
groups which are attempting the fabrication of these structures.

\section{Theoretical method}
\label{sec:theo}

The geometry of the assumed structure is shown in Fig.1a.
The $z$ axis is taken along the DBR growth direction.
We denote by $l_1,l_2$ the thicknesses of the two DBR layers
and by $\varepsilon_1,\varepsilon_2$ their respective dielectric 
constants. The 2D pattern in the $xy$ plane is taken to be 
a triangular lattice of pitch $a$, with a basis in the unit cell
consisting of circular rods of radius $r$ and dielectric
constant $\varepsilon_h$. 
In the present paper we consider only air holes ($\varepsilon_h=1$).
The 3D Bravais lattice is simple hexagonal and the corresponding
Brillouin zone with the main symmetry points is shown in Fig.1b.
The DBR period $L=l_1+l_2$ is taken as a fixed length scale
of the problem, in the sense that the photonic frequencies
are given in dimensionless units $\omega L/(2\pi c)$: indeed,
if such a structure is grown, a single epitaxial DBR can be 
patterned with various lateral periods at fixed $L$.
The 3D photonic lattice is fully determined by the 
ratios $l_1/L$, $a/L$, $r/a$ as well as by the dielectric 
constants $\varepsilon_1,\varepsilon_2,\varepsilon_h$.

In order to calculate the photonic band structure we start 
from the second-order vector equation for the magnetic field,
\begin{equation}
\nabla\times\frac{1}{\varepsilon(\mbox{\bf r})}\nabla\times\mbox{\bf H}
=\frac{\omega^2}{c^2}\mbox{\bf H},
\end{equation}
and use the well-known plane-wave expansion method.
When $N$ reciprocal lattice vectors {\bf G} are kept in the expansion,
the equation is transformed into a $2N\times2N$ matrix eigenvalue
problem which contains the Fourier transform 
$\varepsilon^{-1}_{\mbox{\small {\bf G,G}}'}$
of the inverse dielectric constant $\varepsilon^{-1}(\mbox{\bf r})$.
This is evaluated by the method of Ho, Chan and 
Soukoulis \cite{ho90}, which consists of calculating
the Fourier transform $\varepsilon_{\mbox{\small {\bf G,G}}'}$
of $\varepsilon(\mbox{\bf r})$ and inverting numerically the resulting matrix. 
This procedure yields a rapid numerical convergence when an energy 
cutoff is imposed for the number of reciprocal lattice vectors.

The density of states is defined as
\begin{equation}
  \rho(\omega)=\sum_n\sum_{\mbox{\small\bf k}}
  \delta(\omega-\omega_n(\mbox{\bf k})),
  \label{eq:dos}
\end{equation}
\par

\begin{figure}
\vspace*{3cm}
\resizebox{0.4\textwidth}{!}{%
  \includegraphics{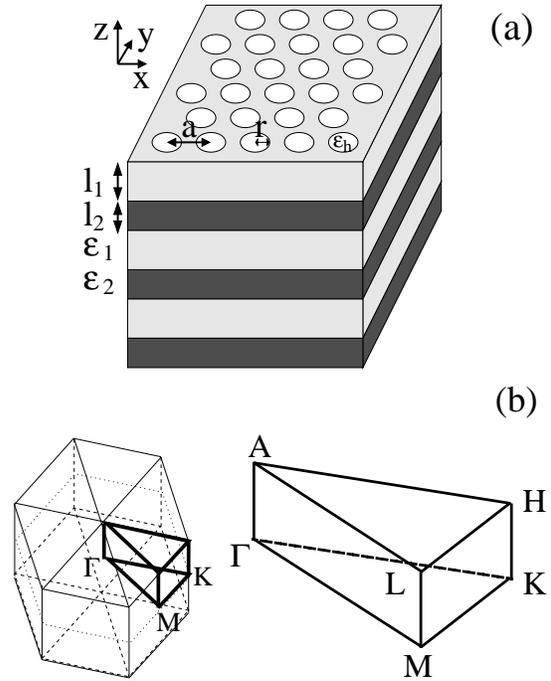}  }\\[1ex]
\caption{(a) Schematic structure of a DBR patterned
with a triangular lattice of circular holes;
(b) the hexagonal Brillouin zone with the irreducible wedge.
The origin of the coordinates in (a) is taken at the center
of a hole in the middle of layer 1.
}
\label{fig:1}
\end{figure}
where $n$ is the index of photonic bands and the sum 
over wavevectors extends over the first Brillouin zone.
The local DOS is similarly defined as
\begin{equation}
  \rho(\omega,\mbox{\bf r})=\sum_n\sum_{\mbox{\small\bf k}}
  |\mbox{\bf E}_{n\mbox{\small\bf k}}(\mbox{\bf r})|^2
  \delta(\omega-\omega_n(\mbox{\bf k})).
  \label{eq:ldos}
\end{equation}
In evaluating Eq.~\ref{eq:ldos} the integral of the squared 
electric field is normalized to the unit cell volume,
therefore DOS and local DOS have the same dimensions 
and can be directly compared.
In order to calculate both quantities we adopt the linear 
tetrahedron method\cite{jepsen71,lehmann72,jepsen84}. 
The DOS is already converged with a mesh of about 
400 $\mbox{\bf k}$-points in the irreducible wedge 
of the Brillouin zone (see Fig.1b).
Calculating the local DOS requires a mesh which spans
the whole Brillouin zone.

We now calculate the Fourier transform of the dielectric tensor,
defined by
\begin{equation}
 \varepsilon_{\mbox{\small {\bf G,G}}'}\equiv 
 \varepsilon(\mbox{\bf G}-\mbox{\bf G}')=
 \frac{1}{V} \int_{\mathrm{cell}} 
 \, \varepsilon(\mbox{\bf r}) 
 \, e^{i(\mbox{\small\bf G}-\mbox{\small {\bf G}}')
 \cdot\mbox{\small\bf r}}
 \, d\mbox{\bf r} ,
 \label{eq:ft}
\end{equation}
for the patterned DBR structure ($V$ is the unit cell volume).
\end{multicols}
\begin{figure*}
\resizebox{.85\textwidth}{!}{%
  \includegraphics{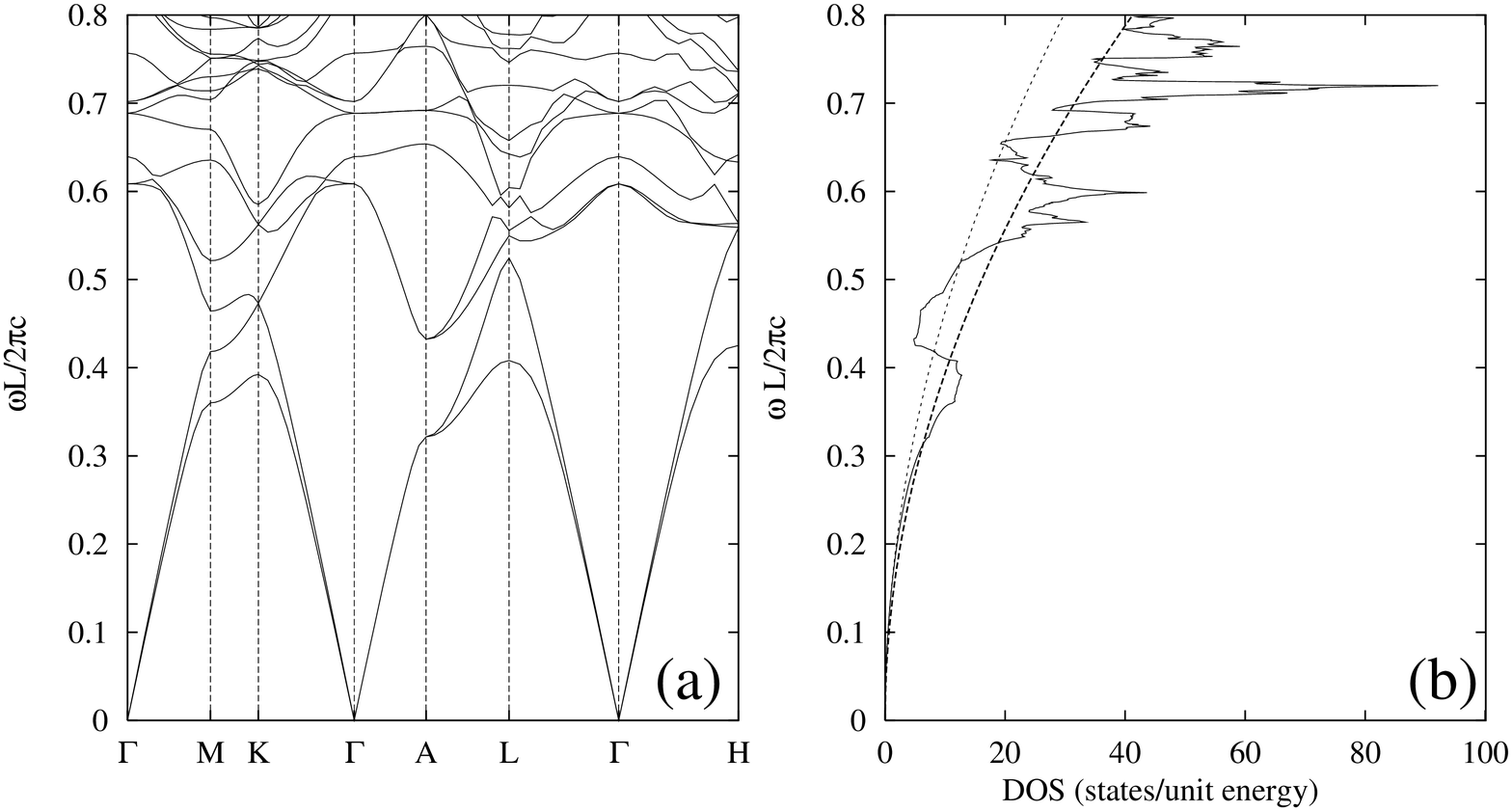} }\\[1ex]
\caption{Bands (a) and density of states (b)
 for a $\lambda/4$ Si/SiO$_2$ multilayer.
 Parameters: $\varepsilon_1=12$, $\varepsilon_2=2$, 
 $l_1/L=0.3$, $a/L=1$, and $r/a=0.45$.
 The dashed line in (b) represents the photon DOS 
 in an average isotropic medium, while the dotted line 
 is the DOS of a uniaxial medium, whose dielectric tensor 
 components are deduced from the photonic bands in the
 long-wavelength limit (see text).
}
\label{fig:2}
\end{figure*}
\begin{multicols}{2}

We write the 3D reciprocal lattice vectors as
$\mbox{\bf G}=({\bf G}_{\parallel},G_z)$,
where ${\bf G}_{\parallel}$ is the projection in the $xy$ plane;
similarly, $\mbox{\bf r}=({\bf r}_{\parallel},z)$.
The unit cell volume is written as $V=AL$, 
where $A$ is the unit cell area for the 2D lattice.
For a general patterning, the Fourier transform can be expressed
in terms of a 2D structure factor
\begin{equation}
 F(\mbox{\bf G}_{\parallel})=
 \frac{1}{A} \int_{\mathrm{diel}}
 \, e^{i\mbox{\small\bf G}_{\parallel}\cdot\mbox{\small\bf r}_{\parallel}}
 \, d\mbox{\bf r}_{\parallel},
\label{strucfac}
\end{equation}
where the integral extends only over the {\em unpatterned} 
(or dielectric) region. Note that $F(\mbox{\bf G}_{\parallel}=0)\equiv f$ 
is the 2D dielectric filling fraction.
For $\mbox{\bf G}=0$, the 3D Fourier transform is simply
the average dielectric constant
\begin{equation}
 \varepsilon(\mbox{\bf G}_{\parallel}=0,G_z=0)=
 (\frac{l_1}{L}\varepsilon_1+\frac{l_2}{L}\varepsilon_2)f+\varepsilon_h(1-f)
 \equiv\varepsilon_{\mathrm{av}}.
 \label{eq:ft1}
\end{equation}
For $\mbox{\bf G}\neq0$, it can be evaluated by subtracting 
the vanishing quantity
$\frac{\varepsilon_h}{V} \int_{\mathrm{cell}}
 \, e^{i\mbox{\small\bf G}\cdot\mbox{\small\bf r}} \, d\mbox{\bf r}$
from Eq.~(\ref{eq:ft}): 
the integrand $\varepsilon(\mbox{\bf r})-\varepsilon_h$
now vanishes in the holes and the integral 
over the dielectric region
yields the 2D structure factor (\ref{strucfac}),
with the result
\begin{equation}
 \varepsilon(\mbox{\bf G}_{\parallel}\neq0,G_z=0)=
(\frac{l_1}{L}\varepsilon_1+\frac{l_2}{L}\varepsilon_2-\varepsilon_h)
 F(\mbox{\bf G}_{\parallel}),
 \label{eq:ft2}
\end{equation}
\begin{equation}
 \varepsilon(\mbox{\bf G}_{\parallel},G_z\neq0)=
 \frac{\sin(G_zl_1/2)}{G_zL/2}
 (\varepsilon_1-\varepsilon_2) F(\mbox{\bf G}_{\parallel}).
 \label{eq:ft3}
\end{equation}
Note that the last expression is valid for any $\mbox{\bf G}_{\parallel}$
($=0$ or $\neq0$).
The above equations (\ref{eq:ft1})-(\ref{eq:ft3}) hold 
for a patterning with a general 2D lattice and holes of arbitrary shape: 
the 2D structure factor can be written as
\begin{equation}
 F(\mbox{\bf G}_{\parallel})=
 \delta_{\mbox{\bf G}_{\parallel},\mbox{\bf 0}}f
 -(1-\delta_{\mbox{\bf G}_{\parallel},\mbox{\bf 0}}) 
 \frac{1}{A} \int_{\mathrm{hole}}
 \, e^{i\mbox{\small\bf G}_{\parallel}\cdot\mbox{\small\bf r}_{\parallel}}
 \, d\mbox{\bf r}_{\parallel}.
\label{general}
\end{equation}
For the present case of a triangular lattice of circular holes,
the unit cell area is $A=\sqrt{3}a^2/2$ and the 2D structure factor becomes
\begin{equation}
F(\mbox{\bf G}_{\parallel}=0)=1-\frac{\pi r^2}{A},
\end{equation}
\begin{equation}
F(\mbox{\bf G}_{\parallel}\neq0)=
-\frac{2\pi}{A}\frac{r}{G_{\parallel}}J_1(G_{\parallel}r),
\end{equation}
where $J_1$ is the Bessel function of first order.

\section{Results}
\label{sec:results}

\subsection{Si/SiO$_2$ system}

We first consider the case of Si/SiO$_2$ multilayers, 
modeled by the dielectric constants $\varepsilon_1=12$
and $\varepsilon_2=2$.
The Si/SiO$_2$ system has the appealing features that the
dielectric contrast is large, implying that a high rejection
ratio in the stop band is obtained with a small number of 
periods, and that the etching of the two materials is a 
technologically well controlled process.
Another high-index contrast system described by similar parameters
is GaAs/Alox (oxidized AlAs) \cite{fiore98}.

In Fig.~2 we show the photonic bands and density of states
in the case of a Si layer width $l_1/L=0.3$: this is close
to the $\lambda/4$ condition for the multilayer (i.e., when
the two layer thicknesses are inversely proportional to their
refractive indices) and it maximizes the 1D gap along the
vertical direction. For the 2D lattice we choose $a/L=1$
and a hole radius $r/a=0.45$: the latter value produces
a full photonic band gap in the pure 2D case, when the refractive
index contrast between dielectric and air is strong enough
\cite{joannopoulos_book,villeneuve92,meade92,padjen94}.
Inspection of Fig.~2a shows that a 1D gap along the $\Gamma$-A
direction is indeed realized around $\omega L/(2\pi c)=0.4$, 
but also that no 2D gap occurs in the 2D projection 
of the Brillouin zone ($\Gamma$-M-K-$\Gamma$ directions).
The reason is that the average dielectric constant of the multilayer
is too low to support a 2D band gap. In order to achieve
a 2D photonic gap, the $\lambda/4$ condition must be abandoned
and multilayers with a higher Si fraction should be considered.

Concerning the photonic density of states, Fig.~2b shows that 
the DOS increases like $\omega^2$ at small frequencies (like
for free photons) but pronounced structures occur when the photonic band
dispersion deviates from linearity.
In this and the following figures for the DOS, the dashed line
represents the photon DOS in a homogeneous and isotropic medium 
with the same average dielectric constant 
$\varepsilon_{\mathrm{av}}$ of the etched DBR (Eq.~\ref{eq:ft1}),
while the dotted line is the DOS of a homogeneous but uniaxial medium, 
whose dielectric tensor components are obtained from the in-plane 
and $\Gamma$-A dispersions at small wavevector \cite{dos}.
The dotted curve agrees with the numerically calculated DOS
in the low-frequency (long-wavelength) limit; the dashed curve
represents a reference DOS of an ``average'' medium, 
and is useful in order to tell whether the DOS of the etched DBR
is reduced or increased with respect to that of the average medium.

In Fig.~3 we show the photonic bands and DOS of three Si/SiO$_2$ 
multilayers with $l_1/L=0.8$ and $a/L=0.8,1.2,1.8$: 
in all these cases the hole radius $r/a=0.45$.
A full 2D band gap in the $\Gamma$-M-K-$\Gamma$ plane can be
recognized in Fig.~3c (around $\omega L/(2\pi c)=0.4$) 
and in Fig.~3e (around $\omega L/(2\pi c)=0.27$),
since the average dielectric constant of the multilayer
is now close to the Si value. In Fig.~3a, the 2D gap should 
coincide with the second-order 1D gap at the $\Gamma$ point:
however, the 2D gap is actually closed by a photonic 
mode which starts at the lower edge of the 1D gap 
at about $\omega L/(2\pi c)=0.51$.
On increasing the ratio $a/L$ the 2D gap opens and decreases 
in energy, and for $a/L=1.8$ (Fig.~3e) it overlaps
the first-order 1D gap at the A point.

In the case shown in Fig.~3e, it would seem that the 1D gap 
along $\Gamma$-A is closed by a photonic mode which starts 
at the lower 2D gap edge around $\omega L/(2\pi c)=0.27$.
However, since this mode is nondegenerate, it does not have
the same symmetry of the electromagnetic field (which belongs
to a twofold degenerate representation for any wavevector {\bf k}
along $\Gamma$-A) and it cannot couple 
to an external beam incident along the $\Gamma$-A direction: this mode 
is ``symmetry uncoupled'' \cite{sakoda_book,robertson94,sakoda95,galli02}
and it cannot have any observable effect on the reflectance 
for a plane wave incident along the multilayer axis.
Indeed, calculations of the optical properties \cite{unpublished}
indicate that the etched DBR still behaves as a 1D reflector
with a well-defined stop band along $\Gamma$-A
and with a reflectivity which tends to unity as 
the number of periods increases.

Concerning the photonic DOS, most features are similar
to those of Fig.~2b already commented:
however it is interesting to observe that a pronounced minimum
occurs in correspondence with the 2D gap along $\Gamma$-M-K-$\Gamma$
(e.g., around $\omega L/(2\pi c)=0.27$ in Fig.~3f).
In the frequency window of the 2D gap the DOS is reduced
not only with respect to the ``average'' DOS (dashed line),
but also in comparison with the DOS of the long wavelength limit
(dotted line). The reduction of the DOS is particularly pronounced
in the cases of Figs.~3 (b) and (f), when the 2D gap
overlaps the 1D gap.
Therefore, a reduction of spontaneous emission rates and a strong 
change of emission pattern is expected under these conditions: 
for the parameters of Fig.~3 (e,f), since spontaneous emission
is suppressed in the $xy$ plane {\em and\/} along the
vertical direction (due again to symmetry mismatch between
the emitted photon and the nondegenerate mode in the gap),
the emission of a point-like dipole will be preferentially 
directed along the diagonals of the 3D structure.

In Fig.~\ref{fig:ldos} we show the local density of states (LDOS)
for the ``optimal'' parameters of Fig.~3(e,f), i.e.,
when the 2D and first-order 1D gap overlap.
The LDOS is evaluated at three different positions along a line 
connecting two neighboring holes in the midplane of layer 1.
The general behavior of the LDOS is similar that of the DOS,
with a $\omega^2$ increase at small frequencies, a reduction
in correspondence with the common 2D-1D gap and pronounced structures
at higher frequencies. As discussed in connection with
Eqs.~\ref{eq:dos},\ref{eq:ldos}, the DOS and LDOS are chosen
to have the same units and can be directly compared, 
or in other words, the spatial average of the LDOS 
over all positions gives again the DOS.
It can be seen that the LDOS is increased over the DOS 
of Fig.~3(f) for a position inside a hole (Fig.~\ref{fig:ldos}a or b),
while it is considerably reduced at the center of the narrow
dielectric region between two holes (Fig.~\ref{fig:ldos}c).
The combined effects of a common 2D-1D gaps and of a change
in position (i.e., of the electric field profile)
results in a sizeable reduction of the LDOS:
at $\mbox{\bf r}=(0,0.5a,0)$ and $\omega L/(2\pi c)=0.28$,
the LDOS is equal to 4.4, compared to about 33
for the average medium. 
Therefore, the numerical results confirm that a strong
reduction of spontaneous emission rates can be achieved
for an emitter embedded in an etched DBR, when 
the parameters of the structure are properly chosen.

\end{multicols}
\begin{figure*}
\resizebox{!}{.85\textheight}{%
  \includegraphics{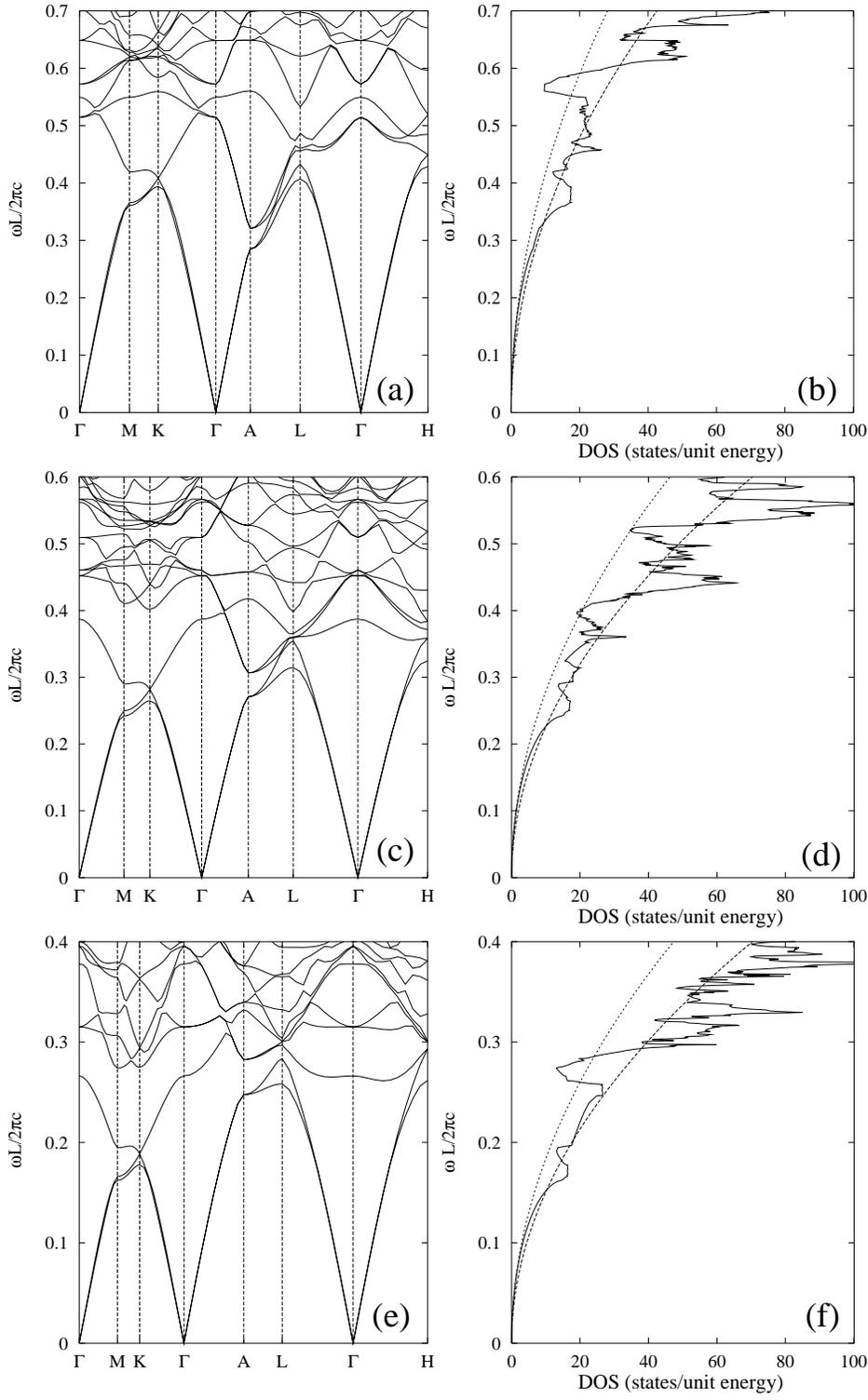} }\\[1ex]
\caption{Bands (left) and density of states (right)
 for Si/SiO$_2$ multilayers. Parameters: $\varepsilon_1=12$,
 $\varepsilon_2=2$, $l_1/L=0.8$, and $r/L=0.45$.
 Figs.~(a,b): $a/L=0.8$, Figs.~(c,d) $a/L=1.2$, Figs.~(e,f) $a/L=1.8$.
 Dashed and dotted lines in the right panels:
 same as in Fig.\ref{fig:2}.
}
\label{fig:3}  
\end{figure*}

\begin{figure*}
\resizebox{0.9\textwidth}{!}{%
 \includegraphics{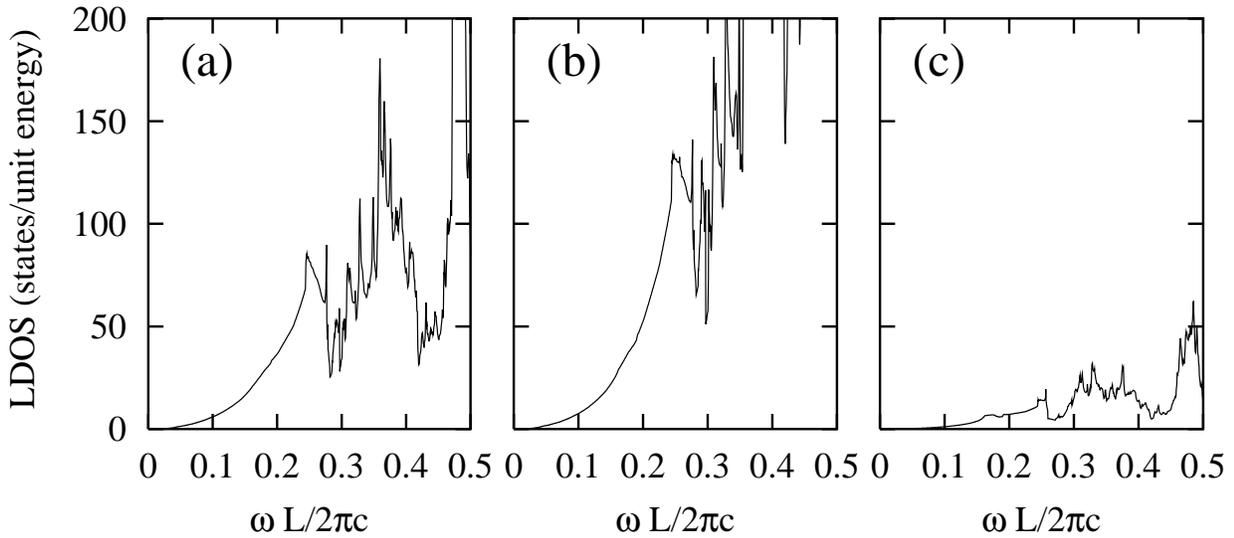}}\\[1ex]
\caption{Local density of states (LDOS) for Si/SiO$_2$ multilayers.
 Parameters: $\varepsilon_1=12$, $\varepsilon_2=2$, 
 $l_1/L=0.8$, $a/L=1.8$, and $r/a=0.45$.
 (a): LDOS at position $(0,0,0)$, i.e., at the center of a hole 
 in the middle of layer 1;
 (b): LDOS at position $(0,0.3a,0)$;
 (c): LDOS at position $(0,0.5a,0)$, i.e., at the center of the dielectric
 vein between two neighboring holes.
 }
\label{fig:ldos}  
\end{figure*}
\begin{multicols}{2}

\subsection{GaAs/AlGaAs system}

\begin{figure*}
\resizebox{0.85\textwidth}{!}{%
  \includegraphics{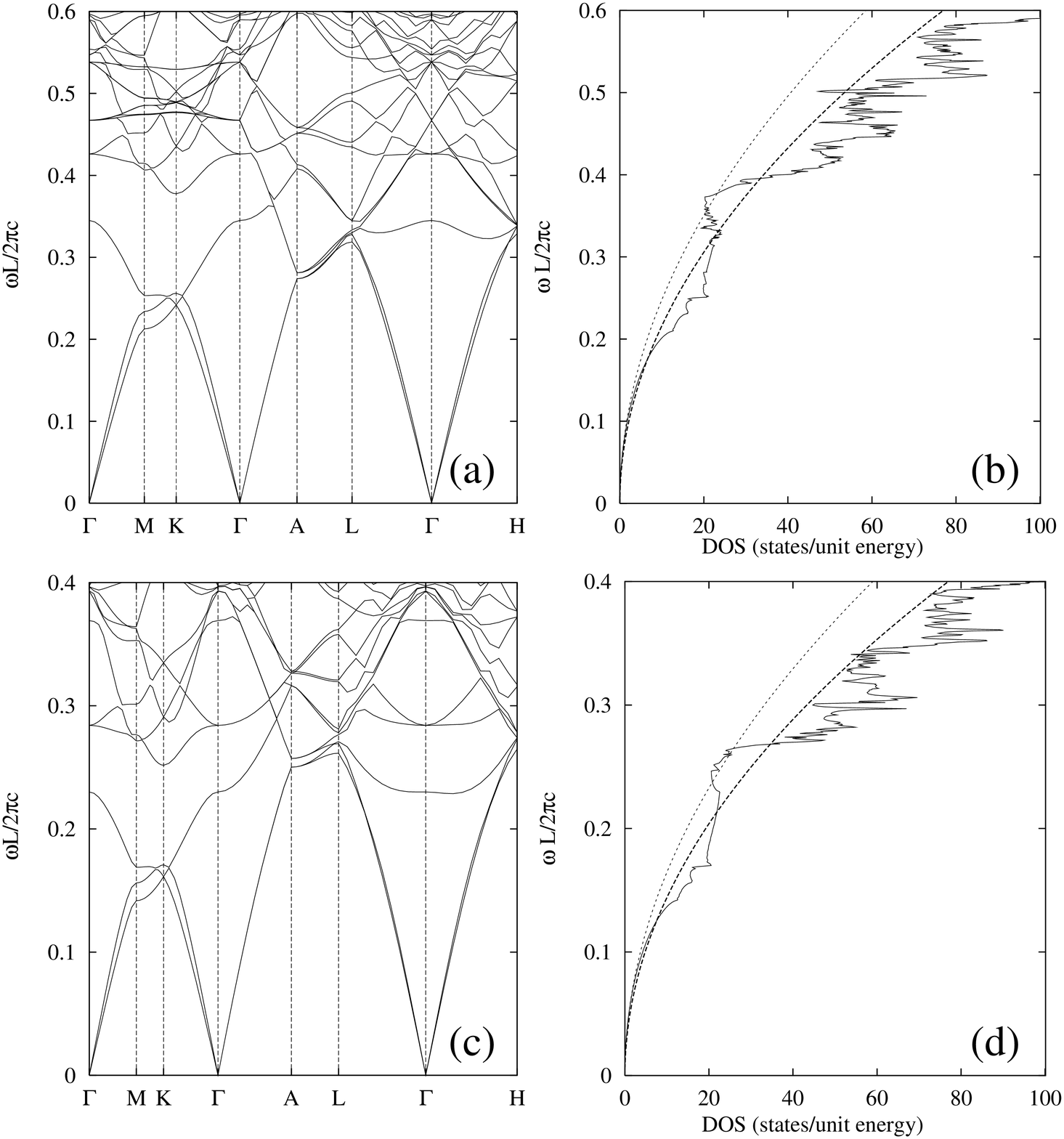} }\\[1ex]
\caption{Bands (left) and density of states (right)
 for GaAs/AlGaAs multilayers.
 Parameters: $\varepsilon_1=11.7$, $\varepsilon_2=10.5$,  
 $l_1/L=0.5$, $r/a=0.45$.
 Figs.~(a,b): $a/L=1.2$, Figs.~(c,d) $a/L=1.8$.
 Dashed and dotted lines in the right panels:
 same as in Fig.\ref{fig:2}.
           }
\label{fig:gaas} 
\end{figure*}

We now consider the case of GaAs/AlGaAs multilayers, 
modeled by the parameters $\varepsilon_1=11.7$ 
and $\varepsilon_2=10.5$: the two values correspond to the
dielectric constants of GaAs and AlGaAs, respectively, 
at 1~eV \cite{palik}. Since the two refractive indices 
are large and close to each other, the photonic energies
in dimensionless units do not depend sensitively 
on the layer thicknesses $l_1,l_2$ and we choose them
to be equal, $l_1=l_2=0.5\,L$: this is also close 
to the $\lambda/4$ condition for the multilayer.
In Fig.~\ref{fig:gaas} we show the photonic bands and DOS 
with these parameters, assuming two different values $a/L=1.2$ 
and 1.8 and taking again a large hole radius $r/a=0.45$.
A 2D gap along $\Gamma$-M-K-$\Gamma$ is formed
(around $\omega L/(2\pi c)=0.36$ in Fig.~\ref{fig:gaas}a,
around $\omega L/(2\pi c)=0.24$ in Fig.~\ref{fig:gaas}c)
and in the case of $a/L=1.8$ it overlaps the small 
first-order 1D gap at the A point.
Like in the case of Fig.~3e, the nondegenerate mode
starting at the lower edge of the 2D gap is not coupled
to a plane wave incident along the $\Gamma$-A direction
and it will not affect the normal-incidence reflectivity.
Concerning the density of states, there is again an increase
like $\omega^2$ at small frequencies followed by several structures. 
A reduction of the DOS occurs in correspondence with the 2D gap,
with a minimum which is somewhat less pronounced than for the 
etched Si/SiO$_2$ multilayers: this is due
to the smaller refractive index contrast between GaAs and AlGaAs.
Changes of spontaneous emission rates and patterns will
also occur in the GaAs/AlGaAs system, but to a lesser extent
than for a high index contrast structure.

\section{Conclusions}
\label{sec:conclusions}

Systematic calculations of photonic bands, total and local DOS
have been undertaken for the hexagonal photonic structure
corresponding to a distributed Bragg reflector patterned
with a triangular lattice of holes.
A few representative examples have been shown for the case of
Si/SiO$_2$ and GaAs/AlGaAs multilayers, focusing on the
most favorable situations for achieving a full 2D photonic gap
in the $\Gamma$-M-K-$\Gamma$ plane.
The parameters for which the 2D gap overlaps the 1D gap
formed along the multilayer axis have been determined.
A nondegenerate mode whose dispersion along $\Gamma$-A
overlaps the 1D gap is symmetry uncoupled from a plane wave
propagating along $\Gamma$-A and it will not affect 
the normal-incidence reflectivity.

The conditions for the overlap of a 2D and a 1D gap
are rather restrictive and require a careful structure design:
the average refractive index of the multilayer has to be large
(thus the layer thicknesses must not obey the $\lambda/4$ condition)
and the air fraction of the 2D lattice must be such that
a full 2D gap develops.
The overlap of the 2D with the first-order 1D gap
is found to occur around $a/L\sim1.8$ for the investigated
structure, which is favourable in view of patterning 
the multilayers with lithography and etching techniques.

The photonic density of states has a pronounced minimum
in correspondence with the common 2D-1D gap, 
particularly for the case of the Si/SiO$_2$ system.
The local DOS depends strongly on the position and is considerably
reduced at the center of the dielectric veins between two holes.
The combined effects of overlapping 2D-1D gaps and of 
the position dependent DOS will yield a strong reduction
of spontaneous emission rates and considerable changes
of emission patterns, with a redistribution of the emission 
mainly along the diagonal direction of the 3D structure.

Specific calculations of the optical properties of etched DBRs
(reflectivity and spontaneous emission) are underway \cite{unpublished}.
The anisotropic structure made of the etched DBR
is concluded to be an interesting possibility 
for the realization of 3D photonic crystals 
where a controlled introduction of defects is possible
and whose spontaneous emission properties can be tailored
to a large extent.

\section*{Acknowledgements}

The authors are grateful to Mario Agio for performing
preliminary reflectivity calculations and for useful discussions.
This work was supported in part by INFM PAIS 2001 project "2DPHOCRY"
and by MIUR through Cofin 2000 program.

\end{multicols}

\begin{thebibliography}{}
\bibitem{dos} The photonic DOS in a uniaxial medium is obtained
by summing the DOS of ordinary and extraordinary modes:
if $\rho_0$ is the free-space DOS and $n_{\parallel}$
($n_{\perp}$) is the refractive index for a polarization
parallel (perpendicular) to the optic axis, the DOS 
is given by $\rho=(n_{\perp}^3+n_{\perp}n_{\parallel}^2)\rho_0$.
\end{thebibliography}
\end{document}